\documentclass[12pt,preprint]{aastex}

\newcommand{\lsun}{L$_\odot$}

\shorttitle{And VIII: M31's New Satellite}
\shortauthors{Morrison et al.}

\received{2003 August 15}
\begin{document}


\title{Andromeda VIII -- a New Tidally Distorted Satellite of M31}


\author{Heather L. Morrison\footnote{Cottrell Scholar of Research
Corporation and NSF CAREER fellow}, Paul Harding, Denise
Hurley-Keller\footnote{NSF Astronomy and Astrophysics Postdoctoral Fellow}}
\affil{Department of Astronomy\footnote{and Department of Physics}, 
Case Western Reserve University, Cleveland OH 44106-7215 
\\ electronic mail: heather@vegemite.cwru.edu,harding@dropbear.cwru.edu, denise@smaug.cwru.edu}

\and

\author{George Jacoby}
\affil{WIYN Observatory, P.O. Box 26732,  Tucson, AZ 85726;
electronic mail: jacoby@wiyn.org}



\begin{abstract}

We report the detection of a new satellite of M31, projected close to
M32. And VIII is tidally distorted, with length $\sim$10 kpc and width
a few kpc. It contains 5--12 planetary nebulae (PNe) and 1--3 globular
clusters, and has a velocity of --204 km/s with respect to M31, some
350 km/s away from M32's velocity.  There are also $\sim 4 \times 10^5$
solar masses of HI, well-separated from the disk, at the same position
and velocity. The satellite has luminosity of 1.2--2.4$\times 10^8$
\lsun, and a central surface brightness of order $\mu_V=24$. Both
these values are typical of Local Group dwarf galaxies. Its surface
brightness is some 6 magnitudes brighter than any of the stellar
streams found in the Milky Way or M31.  The three associated globular
clusters have reddening consistent with foreground reddening from the
Milky Way only, making it likely that the satellite is in front of
M31, unlike the giant tidal stream of Ibata et al.(2001), which is
behind M31 in the SE quadrant. However, the major axis of And VIII is
aligned with the western edge of this giant stream, and we suggest
that its unusual fan shape is caused by superposition of two streams,
the westernmost of which was tidally stripped from And VIII.

\end{abstract}




\section{Introduction}

Tidal streams from disrupting dwarf satellites provide a rich fund of
information about both their progenitor satellite and the dark matter
halo of the parent galaxy \citep[eg][]{kathryn}. In the Milky Way,
studies of the streams of the Sgr dwarf galaxy
\citep[eg][]{igi,ibata_cstars,robbie,majewski} are starting to teach
us about both. The detection of tidal streams around M31
\citep{ibata1,annette,ibata2} has raised the prospect of similar
investigations of the accretion history and mass distribution of the
Andromeda galaxy.

Kinematics of stream stars are vital, as they allow us to constrain
the orbit of the stream and progenitor.  These are more difficult to
obtain in M31 than the Milky Way because, at its distance, velocities
of individual stream red giants require 10m-class telescopes. However,
both PNe and globular clusters are far more luminous and observations
with 4m-class telescopes can easily provide good spectra and
velocities.

Unfortunately, both of these tracers are rare: in a survey reaching
2.5 magnitudes down the PNLF, only one PN is detected per $3 \times
10^7$ \lsun\ --- the estimated luminosity of the {\it entire} giant
stream found in the SE region of M31 \citep{ciardullo,ibata1}. Only
the two most luminous dSph galaxies in the Milky Way have PNe: the Sgr
dwarf has two and Fornax one \citep{sgrpn}.  Only objects with larger
luminosity such as M32 will have significant numbers of PN associated
\citep[][~found 15]{nolford}. We expect any PNe 
associated with the M31 streams to be from strong density enhancements
such as the still-bound cores of
accreting satellites.  dSph galaxies of this luminosity also have only a few
globular clusters  \citep{marioaraa}.

There are now high-quality velocities available for over 300 globular
clusters and 135 PNe in M31.  The globular clusters have velocity
errors of 12 km/s \citep{perrett} and the PNe, observed in a new
survey which reaches out to 20 kpc from the center, have even more
accurate velocities with typical errors of 5 km/s \citep{denise}. Such
velocity precision allows, for the first time, sensitive searches both
for disk kinematics \citep{diskglobs} and also velocity
substructure. In this Letter we report on a group of PNe and globular
clusters close to the projected position of M32 which have a very low
velocity dispersion and a mean velocity which differs by 350 km/s from
M32's. There are also two small HI clouds at similar position and
velocity.  We suggest that they belong to a previously undiscovered
satellite of M31's, with luminosity of order 10$^8$ \lsun, which is
currently undergoing tidal disruption.

\section{PN, Cluster and HI Kinematics}

\begin{figure}

\includegraphics[scale=0.65,angle=270]{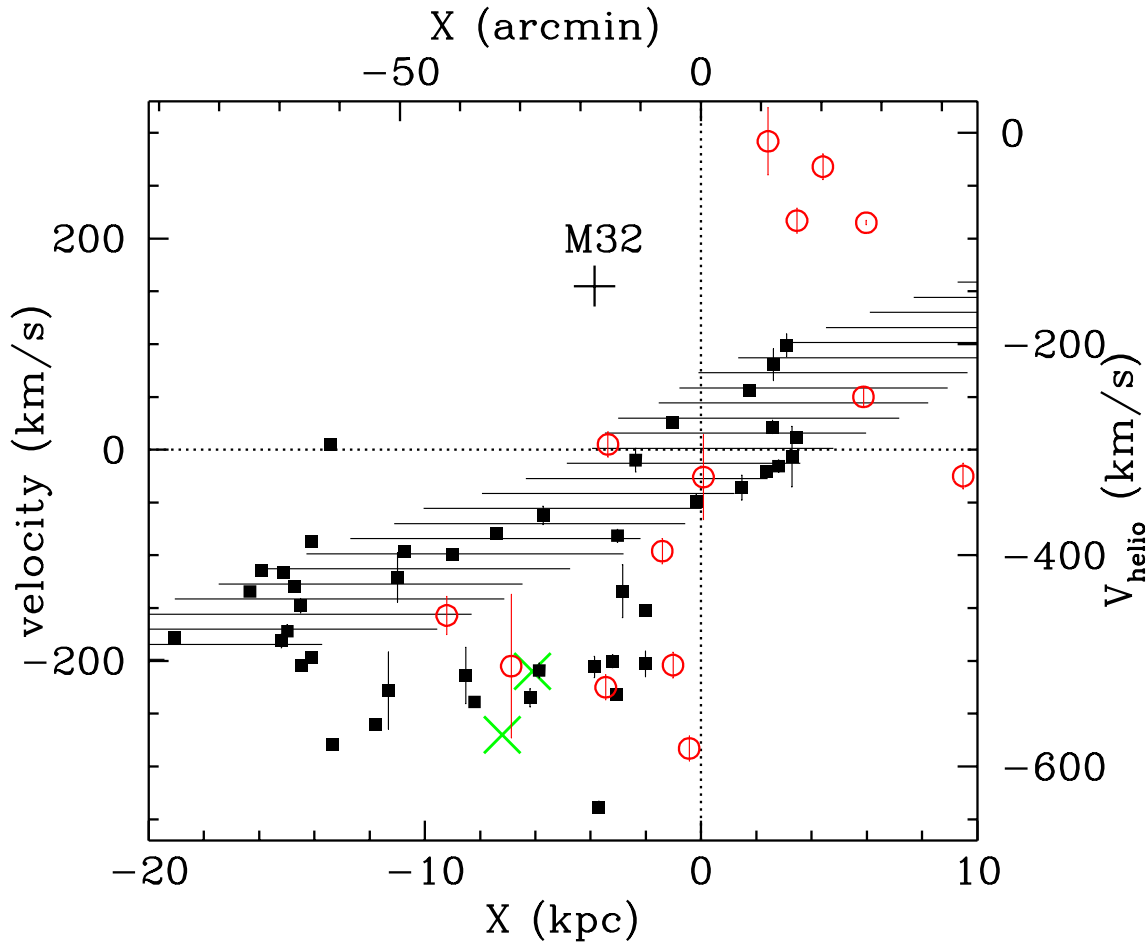}        

\caption{Kinematics of PNe and globular clusters in a strip parallel
to the major axis with Y=--4 to --7 kpc. Velocities are with respect
to M31, and are plotted against distance along the major axis.  The
shaded region shows the expected range of velocity of disk objects, from 100
simulations of disk kinematics as described in \citet{diskglobs}.  PNe
are plotted as black filled squares, globular clusters as open circles
(red in the online edition), HI
detections as crosses (green in the online edition). While more than half of the PNe in this
region show velocities typical of disk objects, there is a narrow
velocity feature stretching over almost 10 kpc with mean velocity
--204 km/s which contains 5--12 PNe and 1--3 globular clusters.
Although M32 is located close to this region on the sky, its velocity
is $\sim$350 km/s away.
\label{streamcore}} 
\end{figure}

To search for substructure in regions projected on the disk, we need
to recognize the disk's kinematical signature.  This is particularly
clear when narrow strips parallel to the major axis are plotted with
velocity vs. X \citep[distance along the major
axis:][]{brinks,diskglobs}.  In Figure \ref{streamcore} we show the
kinematics of PN and globular clusters in a strip with Y (distance
parallel to the minor axis) from --4 to --7 kpc. These Y distances
deproject to distances of 18 to 31 kpc (3 to 6 disk scale lengths)
from the major axis.  Thin disk objects show a diagonal line in this
diagram because much of their circular velocity is projected away from
the line of sight. This region is shown shaded in the Figure.

While more than half of the PNe in this region have disk kinematics,
we also see a narrow velocity feature stretching over almost 10 kpc
with mean velocity $\sim$ --200 km/s with respect to M31 (--500 km/s
heliocentric). It contains 5--12 PNe and 1--3 globular clusters.
Although M32 is located close to this position on the sky, its
velocity is $\sim$350 km/s away. There is also some
evidence that the feature extends out to X=14 kpc, showing a velocity
gradient. However, we restrict our discussion here to the 5 PNe whose
velocities are, within their individual 1-$\sigma$ errors, identical
to --204 km/s, and the three globular clusters whose velocities agree
with this velocity within their 2-$\sigma$ errors.

\begin{figure}

\includegraphics[scale=0.65]{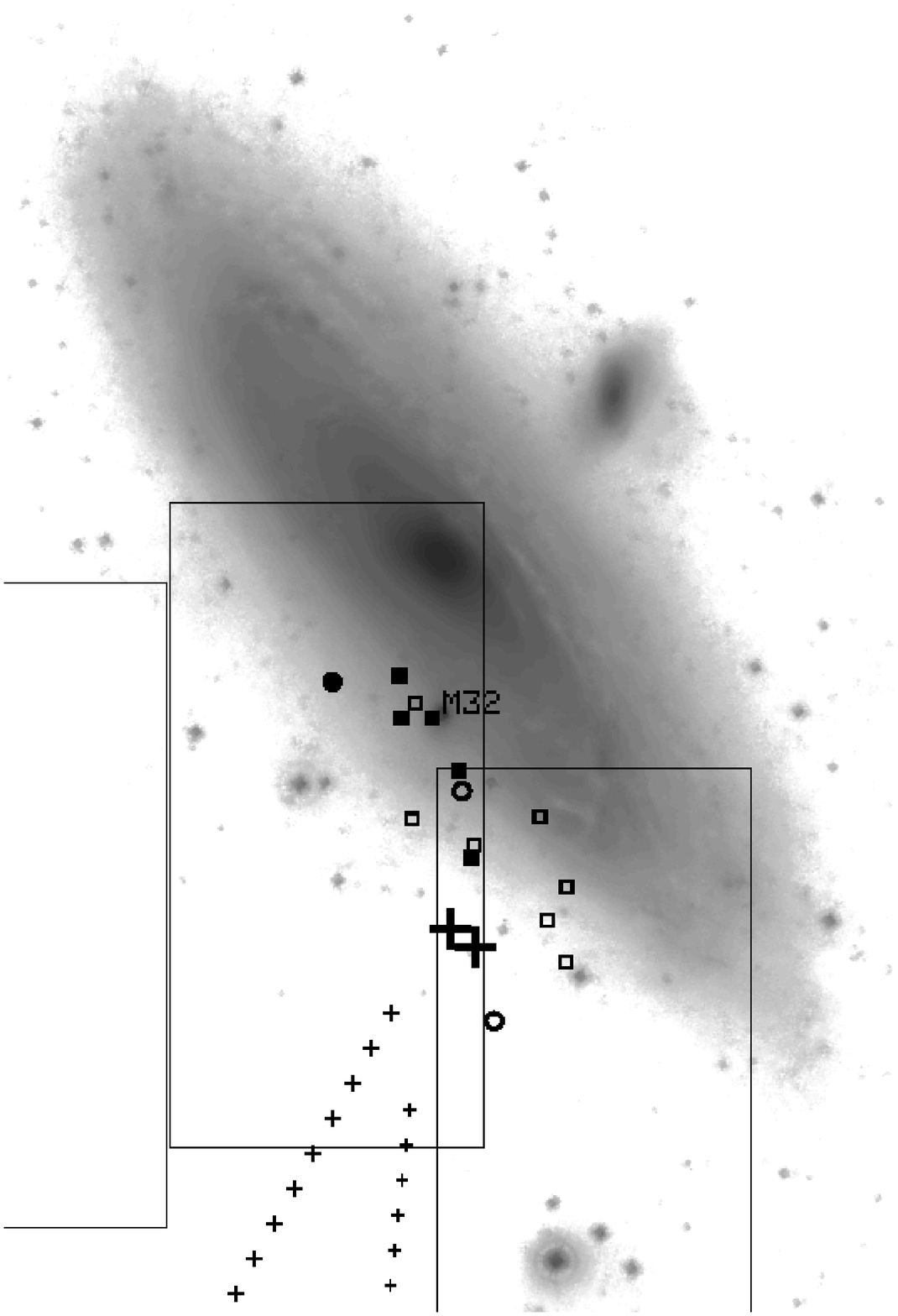}        

\caption{Location of objects in the narrow velocity feature, shown
  overlaid on an image of M31 taken on the Burrell Schmidt telescope,
  kindly made available by Rene Walterbos. N is up, E is to the
  left. The limits of our Schmidt survey for the PNe are shown with
  solid black lines.  Circles are globular clusters (red in the online
  edition), squares PNe (blue in the online edition), crosses HI
  detections (green in the online edition).  Solid squares and circles
  are objects which differ by less than 5 km/s from the mean stream
  velocity of --204 km/s, while open symbols denote objects with
  velocity between --200 and --300 km/s, which may be members of the
  feature if it has a velocity gradient.  The edges of the giant
  stream of of \citet{ibata1} are marked with black crosses. The
  stream appears to fan out in its southern regions.  The densest part
  of the stream (its eastern edge) is shown with larger plus signs,
  while the western edge is shown with the near-vertical line of
  smaller plus signs. It is intriguing that the western edge is
  roughly aligned with the body of And VIII.
\label{xyplot}} 
\end{figure}

Figure \ref{xyplot} shows the location of the objects in this velocity
feature. It can be seen that they trace out a very
elongated structure about 10 by 2 kpc in size.  In Tables
\ref{pnmembers} and \ref{gcmembers} we give velocities for all
possible members of this structure, and metallicity and E(B--V)
estimates for the globular clusters.  Data are from \citet{denise} for
the PNe and from \citet{perrett,barmby00} for the globular
clusters. We use the naming convention of \citet{barmby00} for the
globular clusters.

\begin{deluxetable}{lrrrrrr}

\tablecolumns{7}
\tablewidth{0pt}
\tablecaption{Likely PN members of Andromeda VIII \label{pnmembers}}
\tablehead{
\colhead{ID}  & \colhead{RA(2000)} & \colhead{Dec(2000)} & \colhead{X (arcmin)}  &
\colhead{Y (arcmin)} & \colhead{Velocity (km/s)} & 
\colhead{Error}  } 

\startdata

HKPN 97  & 0:43:20.63  & 40:57:44.5 &  --2.015  &   --4.138 &  --502.8   &  12  \\
HKPN 94 & 0:43:15.18  & 40:51:12.7 &  --3.202  &   --4.993 &  --500.3   &  6   \\
HKPN 81 & 0:42:49.84  & 40:51:10.6 &  --3.864  &   --4.149 &  --505.7   &   10 \\
HKPN 75 & 0:42:22.82  & 40:43:06.6 &  --5.860  &   --4.521 &  --509.1   &  7   \\
HKPN 71& 0:42:04.07  & 40:29:37.7 &  --8.515  &   --6.024 &  --513.8   & 27   \\
	      	     	   	              	 	    
HKPN 51 & 0:40:44.02  & 40:25:08.7 &  --11.32  &   --4.032 &  --528.2   & 37   \\
	      	     	   	              	 	    
HKPN 91 & 0:43:05.26  & 40:53:34.1 &  --3.081  &   --4.286 &  --532.3   &  5   \\
HKPN 87& 0:42:56.18  & 40:35:40.4 &  --6.188  &   --6.821 &  --534.8   &  9   \\
HKPN 70 & 0:42:03.33  & 40:31:39.8 &  --8.207  &   --5.677 &    --539.0   &    4 \\
HKPN 62& 0:41:12.34  & 40:35:58.2 &  --8.840  &   --3.281 &  --528.4   &   10 \\
     	           	   	          	    	          
HKPN 57& 0:40:56.88  & 40:20:00.9 &  --11.79  &   --5.253 &  --560.5   &  4   \\
HKPN 49& 0:40:37.96  & 40:13:35.5 &  --13.35  &   --5.645 &  --578.9   &  3   \\

\enddata
\end{deluxetable}

\begin{deluxetable}{lrrrrrrrr}

\tablecolumns{9}
\tablewidth{0pt}
\tablecaption{Likely globular cluster members of And VIII \label{gcmembers}}
\tablehead{
\colhead{ID}  & \colhead{X (arcmin)}  &
\colhead{Y (arcmin)} & \colhead{Velocity (km/s)} & 
\colhead{Error} & \colhead{[Fe/H]} & 
\colhead{Error} & \colhead{E(B--V)} & \colhead{Error} 
}

\startdata

B219-S271 & --4.65 &   --25.40 &  --504 &   12 &   --0.73&
   0.53 &    0.13&  0.03 \\

B176-S227 & --15.84 &--23.49 &  --525 &   12 &   --1.60 &   0.10 &
0.04 & 0.04 \\

B85-S147 & --31.52 & --18.44 &  --505 &   68 &   --1.83  &  0.40 &
 0.15 &  0.07  \\

\enddata
\end{deluxetable}

The HI survey of \citet{brinks} shows two small HI features
well-isolated from the disk HI, at similar position and velocity: they
are shown in Figures \ref{streamcore} and \ref{xyplot}. The mass of HI
associated with each feature is of order $2 \times 10^5$ solar
masses. These HI detections have been confirmed by a new survey of M31
in HI \citep{braun1} which reaches to much lower column density:
\citet{braun2} find a giant HI tail extending tens of kpc away from
the M31 disk to the SE, most apparent at a heliocentric velocity of
--500 km/s.

\section{Discussion}

It is remarkable to find such a tight velocity feature in not only PNe
and globular clusters, but also in HI. Because of the short
evolutionary time of PNe, this feature has a substantial stellar
luminosity, which we calculate below.

It is clear from Figure \ref{streamcore} that it is not associated
with M31's thin disk. Although M31 is known to have a substantial disk
warp, the closeness to the minor axis means that any warp stars would
also have most of their velocity perpendicular to the line of sight.
The probability of five PNe with such low velocity dispersion being
drawn from a smooth, dynamically hot population such as the bulge is
very small\footnote{To make a rough estimate of how likely this is, we
simulated 10,000 samples of 15 non-disk PN from a population with
velocity dispersion 80 km/s. Only 0.2\% of these samples had 5 objects
within 10 km/s. The spatial association would make it less likely
still.}.  The heliocentric velocity of the feature (--504 km/s) also
makes any association with the Milky Way halo unlikely.  We conclude
that the feature is either a tidal stream or a hitherto undiscovered
satellite of M31's, whose projection close to M31 has made it
invisible until velocity data were available.

What is its luminosity?  We use the number of PN per unit luminosity
of \citet{ciardullo} and the limiting magnitude of our survey in this
area \citep[2.5 magnitudes down the PNLF,][]{denise}, and find a
corresponding total luminosity of $1.5 \times 10^8$\lsun\ and an
absolute magnitude of $M_V$=--15.6. (We obtain a similar luminosity by
scaling by the number of PN detected in M32 in our survey.)  This is a factor of $5
\pm 2$ times the total luminosity of the giant star stream in the regions
detected by \citet{ibata1}, who estimate a luminosity of $3 \times
10^7$ \lsun\, an absolute magnitude of $M_V$=--14 and an average
surface brightness of $\mu_V=30$ mag/arcsec$^2$ for the regions of the
giant stream they can detect in their star counts.

The surface brightness associated with this feature is $\mu_V\sim
23.8$ mag/arcsec$^2$, assuming that its central 4 PNe are contained in
a region of 12 kpc$^2$. This is typical for Local Group dwarf
galaxies, which have central surface brightness values ranging from
$\mu_V=22.1$ to 26.5 mag/arcsec$^2$ \citep{marioaraa}. 
This estimate of the central surface brightness explains why the
feature has not been discovered with imaging surveys. It subtends
almost a degree on the sky, and has significantly lower surface
brighness than the M31 disk in this region. The disk has $\mu_V=20.9$
at a distance of
7 kpc out on the SE minor axis, and the surface photometry of \citet{wk88}
(kindly made available by Rene Walterbos) shows that the galaxy surface
brightness is roughly constant between the minor axis and X=--10 kpc
over the region occupied by this feature.

Although the elongated shape of the feature could indicate that we are
observing a portion of a star stream rather than a tidally distorted
but still bound satellite, its luminosity  places it at the bright end of the existing dSph {\it
satellites} in the Local Group. Its surface brightness is some 6
magnitudes brighter than the giant stream discovered by
\citet{ibata1} and the star streams discovered in the Milky Way
halo. In addition, it appears to stop abruptly near the minor axis:
although we have detected a number of PNe and globular clusters at
similar Y values and positive X, none of these objects have velocities
near --500 km/s. We conclude that this is a still-bound satellite of
M31's which is tidally distorted but not completely disrupted, and we
christen it And VIII.

Its large velocity difference with M32 makes it unlikely that they are
physically associated.  

The reddening of the globular clusters provides an interesting
constraint. All three clusters have low values of E(B--V) \citep[0.13,
0.15 and 0.04 respectively,][]{barmby00}, which are consistent with
the foreground reddening from the Milky Way
\citep[E(B--V)=0.08,][]{burstein}. If the stream were located on the
far side of M31, we would expect to see a significantly higher
reddening of these globular clusters. \citet{ford78} estimate that
there is 0.5 mag. of reddening due to M31's disk at the position of
M32. We examined the globular cluster reddening values classified as
``good'' for clusters with $|Y|>4$ kpc from \citet{barmby00} , and
found reddening values ranging from E(B-V) = 0.04 to 0.40. Since we
would expect the globular clusters to be scattered both in front and
behind the galaxy, this suggests that the reddening due to M31's disk
is at least as high as E(B-V)=0.40.  We conclude that the satellite is
likely located on the near side of M31. The outer regions of the giant
stellar stream in the SE quadrant are on the far side of M31
\citep{ibata2}, so it is unlikely that And VIII is part of it. This is
also consistent with the fact that it is elongated roughly at right
angles to the giant stream.  However, it is intriguing to note that
the giant stream appears to fan out in its southern portions (see, for
example, Figure 2 of \citet{annette}).  The westernmost boundary of
this fan-shaped feature is roughly aligned with the And VIII major
axis: further imaging and PN detections in this region of the stream
will allow us to test whether it is formed by the disruption of And
VIII.

The position of the And VIII HI detections is intriguing because they
seem slightly offset from the stellar locus of And VIII. While the
survey of \citet{braun1} will provide much more accurate positional
information for the HI associated with And VIII, we note that ram
pressure stripping from hot gas in M31's halo might cause the HI to
follow a different locus from the stars.

Is there any more substructure of such relatively high surface
brightness seen in the regions we surveyed?  In this paper we have
used the very low velocity dispersion of the PN and globular clusters
plus their spatial clustering to infer that they belong to a separate
satellite of M31's. We examined the regions covered by our PN survey
(an area bounded roughly by X=--20 to +10 kpc and Y=0 to --20 kpc) for
other substructure and found no strong evidence for other
substructure.  This constrains the existence of other satellites in
the large area covered by our survey.

\section{Summary}

Using high-precision velocities of both PNe and globular clusters
in a region near M32, we have located a feature with very low velocity
dispersion (unresolved within our errors) which contains 5--12 PNe,
1--3 globular clusters and two HI clouds. Its velocity (--204 km/s
with respect to M31) differs from M32's by some 350 km/s. It is
projected on the outer disk of M31,
which explains why it has not been detected before.

It shows an elongated shape (size $\sim$10 by 2 kpc), covers almost a
degree on the sky, and has a total luminosity of $\sim 10^8$ \lsun\
and a mean surface brightness of $\mu_V \sim 24$ mag/arcsec$^2$. These
values are typical of the Local Group dwarf galaxies, and some 6
magnitudes brighter than the tidal streams discovered in the Milky Way
and M31. We conclude that it is a tidally disrupting dwarf satellite
of M31.  There are two HI features associated with And VIII which
contain of order $4 \times 10^5$ solar masses, about 0.1\% of its
total luminous mass. And VIII appears to be a dSph galaxy with a small
amount of associated HI.

All three globular clusters have E(B--V) values consistent with
foreground reddening from the Milky Way, so it is likely that And
VIII, like M32, is in front of M31. The giant stellar stream of
\citet{ibata1} is behind M31 in this region, so And VIII, if it is on
the same orbit as this stream, must be on a very different part of
it. The giant stream displays an unusual fan shape, and the
westernmost edge of the fan aligns roughly with the major axis of And
VIII. It is possible that this feature is actually two streams
superposed on the line of sight, with the westernmost stream tidally
stripped from And VIII.

Velocity measurements of stream stars will soon be available,
allowing us to constrain the orbits of star streams in M31,
investigate their connection with And VIII, and use them to study
M31's dark halo.



\acknowledgments

HLM acknowledges the support of NSF CAREER grant AST-9624542, DHK an
NSF Astronomy and Astrophysics Postdoctoral Fellowship AST-0104455. We
would like to thank Robert Braun for generously sharing his result on
the HI stream in advance of publication, and an anonymous referee
whose suggestions substantially improved the paper.  This discovery
was made while HLM and PH were visiting Mt Stromlo, RSAA, ANU. We
thank Mt Stromlo for its hospitality and for helpful conversations
with many of its astronomers, particularly Gary Da Costa, Penny
Sackett and Frank Briggs.

\end{document}